# Dependence of nucleotide physical properties on their placement in codons and determinative degree

BEREZHNOY A.Y. [1], DUPLIJ S.A. [2]
([1]NSC Kharkov Institute of Physics and Technology, Kharkov 61108, Ukraine)
([2]Department of Physics and Technology, Kharkov National University Svoboda Sq.4, Kharkov 61077, Ukraine)
E-mail: yuberezhnoy@pht.univer.kharkov.ua; Steven.A.Duplij@univer.kharkov.ua

**Abstract:** Various physical properties such as dipole moment, heat of formation and energy of the most stable formation of nucleotides and bases were calculated by PM3 (modified neglect of diatomic overlap, parametric method number 3) and AM1 (austin model 1) methods. As distinct from previous calculations, for nucleotides the interaction with neighbours is taken into account up to gradient of convergence equaling 1. The dependencies of these variables from the place in the codon and the determinative degree were obtained. The difference of these variables for codons and anticodons is shown.

## INTRODUCTION

It is well-known from the genetic code structure that there exists a connection between codons and physical and chemical amino acid properties (Lewin, 1983). Codon properties in turn are defined by physical properties of the nucleotides from which they are compounded (Ratner, 1985; 2000). The most important properties of nucleotides as complex three-dimensional molecules are their dipole moment, heat of formation and energy of the most stable conformation (Schneider and Berman, 1995; Zheltovsky *et al*., 1989; Govorun *et al*., 1992). This work is devoted to theoretical calculation of these parameters depending on the nucleotide position in the codon and on the characteristics descriptions of different bases specificity−the nucleotides determinative degree introduced in (Duplij and Duplij, 2000; 2001; Duplij *et al*., 2000).

Various physical properties, such as total energy, heat of formation, dipole moment and ionization potential of canonical isolated nucleotides were calculated in (Zheltovsky *et al*., 1989; Govorun *et al*., 1992) by AM1 (austin model 1) method and in (Sponer *et al*., 1996) by MP2 (second-order Moller-Plesset perturbational method). Here we use PM3 (modified neglect of diatomic overlap, parametric method number 3) method which is a modified AM1 method with many more parameters (Stewart, 1990; Dewar *et al*., 1985).

## GENETIC CODE AND DETERMINATIVE DEGREE OF NUCLEOTIDES

It is known that the bases in the genetic code triplets play different roles related to amino acid determination. For example, first base doublet in a codon determines certain amino acid's formation to a greater degree than third base doublet in codon. Hence half of all triplets (32 codons) have full degeneration by third base doublet, so that an amino acid is entirely specified by the first two nucleotides independently of the third one (Lewin, 1983; Singer and Berg, 1991). Nearly two thirds of all DNA bases have almost constant properties in every organ-

ism−those are the bases being in the first or second positions in a triplet. Sixteen of possible doublets may be described as two octets. The first eight doublets ("strong") code amino acids independently of the third codon's nucleotide base and another eight doublets ("weak") determine the codon by third nucleotide base. So it is possible to arrange nucleotides in decreasing order of their ability of amino acid one-to-one determination in such a way: C, G, T, A (Rumer, 1968). In the works (Duplij and Duplij, 2000; Duplij *et al.*, 2000) an abstract characteristics of nucleotides−the determinative degree ($d_x$)−was introduced and used for numerical description of the "strength" of nucleotides. It was proposed so as shown as follows:

| Pyrimidine | Purine | Pyrimidine | Purine |
|---|---|---|---|
| C | G | T | A |
| Very strong | Strong | Weak | Very weak |
| $d_C$=4 | $d_G$=3 | $d_T$=2 | $d_A$=1 |

It allows us to follow from the qualitative description of the genetic code structure (relative to ability of coding amino acids) to its quantitative description. We suppose that the first approximation determinative degree of codons is additive, that is for doublet $x$, $y$ we have $d_{xy}=d_x+d_y$. So the rhombic doublets structure can be presented as follows:

| | | | | | | | |
|---|---|---|---|---|---|---|---|
| | | | CC | | | | =8 |
| | | GC | | CG | | | =7 Strong |
| | CU | | GG | | CU | | =6 |
| AC | | UG | | GU | | CA | =5 Middle |
| | AG | | UU | | GA | | =4 |
| | | AU | | UA | | | =3 Weak |
| | | | AA | | | | =2 |

which corresponds to the rhombic version of the genetic dictionary (Karasev and Sorokin, 1997; Karasev, 1976), and is related to the "strength" of every doublet so that horizontal rows of this structure consist of doublets with equal "strength". The additivity proposition for triplets gives the possibility to calculate their "strength" by the formula $d_{xyz}=d_x+d_y+d_z$ and to examine the symmetry of the cubic codon matrix.

Next we deal with physical properties of nucleotides depending on the abstract characteristics−their determinative degree $d$.

METHODS OF CALCULATIONS AM1, PM3

AM1 is a modified MNDO (modified neglect of differential overlap method) method. PM3 is a reparametrization of AM1 which is based on the ignoring of diatomic differential overlap (NDDO) approximation (Stewart, 1990). NDDO retains all one-center differential overlap terms when coulomb and exchange integrals are computed. PM3 differs from AM1 only in the values of the parameters (Dewar *et al.*, 1985). The parameters for PM3 were derived by comparing a much larger number and wider variety of experimental versus computed molecular properties. The elements of the Fock matrix based on the NDDO approximation are described below. When orbitals $\phi_\mu$ and $\phi_\nu$ are on different centers, the off-diagonal elements of the Fock matrix are

$$F_{\mu\nu}^{\alpha} = H_{\mu\nu} - \sum_{\lambda}^{A}\sum_{\sigma}^{B} P_{\lambda\sigma}^{\alpha}\left(\mu\lambda \mid \nu\sigma\right),$$

where $\mu$ on $A$, $\nu$ on $B$, $A\neq B$, and when $\phi_\mu$ and $\phi_\nu$ are different atomic orbitals but on the same center, then off-diagonal elements of the Fock matrix are

$$F_{\mu\nu}^{\alpha} = H_{\mu\nu} + P_{\mu\nu}^{\alpha}[3(\mu\nu \mid \mu\nu) - (\mu\mu \mid \nu\nu)] + \sum_{B}\sum_{\sigma}^{B}\sum_{\lambda}^{B} P_{\lambda\sigma}^{\alpha+\beta}(\mu\nu \mid \lambda\sigma).$$

where $\lambda$ is index of orbitals on atom A and $\sigma$ is index of orbitals on atom B , and α and β desribe two different sets of spatial molecular orbitals−those that hold electrons with spin up and those that hold electrons with spin down, respectively.

The diagonal elements of the Fock matrix are

$$F_{\mu\mu}^{\alpha} = \sum_{\nu}^{A}\left[P_{\nu\nu}^{\alpha+\beta}(\mu\mu \mid \nu\nu) - P_{\nu\nu}^{\alpha}(\mu\mu \mid \nu\nu)\right] + \sum_{B}\sum_{\sigma}^{B}\sum_{\lambda}^{B} P_{\lambda\sigma}^{\alpha+\beta}(\mu\mu \mid \lambda\sigma) + H_{\mu\mu},$$

where $\boldsymbol{P}^{\alpha+\beta}=\boldsymbol{P}^{\alpha}+\boldsymbol{P}^{\beta}$ and $\boldsymbol{P}$ is the effective number of electrons occupying the atomic orbital.

The terms involved in the above equations are described below.

**Two-center two-electron integrals**

The MNDO method has 22 unique two-center two-electron integrals for each pair of heavy

(non-hydrogen) atoms in their local atomic frame. With exception of integral 22, $(p_p p'_p \mid p_p p'_p)$, where the index $p=\sigma$, $\pi$, all other integrals can be computed a priory without loss of rotational invariance. That is, no integral depends on the value of another integral, except for the last one. It can be shown that

$$(p_\pi p'_\pi \mid p_\pi p'_\pi) = \frac{1}{2}(p_\pi p_\pi \mid p_\pi p_\pi) - (p_\pi p_\pi \mid p'_\pi p'_\pi). \quad (1)$$

The two-center two-electron repulsion integrals $\mu\nu|\lambda\sigma$ represent the energy of interaction between the charge distributions ($\phi_\mu\phi_\nu$) in atom $A$ and ($\phi_\lambda\phi_\sigma$) in atom $B$. Because the repulsion interaction energy of two point charges is inversely proportional to the distance separating the two charges, for example the ($ss|ss$) two-centered two-electron integral can be represented by:

$$(ss \mid ss) = \frac{1}{\left((R + c_A + c_B)^2 + \frac{1}{4}(1/A_A + 1/A_B)^2\right)^{1/2}}, \quad (2)$$

where $A_A$, $A_B$ are parameters and $c_A$, $c_B$ are distances of the multipole charges from their respective nuclei $A$ and $B$ (a simple function of the atomic orbital type). There are some boundary conditions which can be used to fix parameters $A_A$ and $A_B$. For example, when the distance ($R$) between nucleus $A$ and nucleus $B$ approaches to zero, i.e., $R_{AB}\rightarrow 0$, the value of the two-center two-electron integral should approach to the corresponding monocentric integral. The MNDO nomenclature for these monocentric integrals is,

$$G_{ss}=(ss|ss),\ G_{sp}=(ss|pp),\ H_{sp}=(sp|sp)$$
$$G_{pp}=(pp|pp),\ G_{p2}=(pp|p'p'),\ H_{pp'}=(pp'|pp') \quad (3)$$

Using the above asymptotic forms of the two-center two-electron integrals, the parameters $A_A$ and $A_B$ can be derived. One uses AM to represent the parameter $A$ obtained via $G_{ss}$, AD to represent the parameter $A$ obtained via $H_{sp}$, and AQ to represent the parameter $A$ obtained from $H_{pp'}$. Therefore, for example the two-center two-electron integral ($ss|ss$) can be written as

$$(ss \mid ss) = \frac{1}{\left(R_{AB}^2 + \frac{1}{2}\left(1/AM_A + 1/AM_B\right)^2\right)^{1/2}}. \quad (4)$$

**One-center one–electron integral $H_{\mu\mu}$**

The diagonal elements of the core Hamiltonian represent the energy of an electron in an atomic orbital of the corresponding atom $U_{\mu\mu}$ plus the attraction of an electron in the atomic orbital (one atom $A$) for the other nuclei (atoms $B$). In the MNDO method, the one-center one-electron integral $H_{\mu\mu}$ is given by:

$$H_{\mu\mu} = U_{\mu\mu} - \sum_{B\neq A} V_{AB}, \quad (5)$$

where $V_{AB}$ represents attraction of an electron in an atomic orbital of the corresponding atom (on atom $A$) for the other nuclei (atoms $B$), and it can be approximated in a similar way as before

$$V_{AB}=Z_B(\mu\mu|ss) \quad (6)$$

where ($\mu\mu|ss$) is the MNDO generalization of the MINDO3 (modified intermediate neglect of differential overlap method) two-electron integral $\gamma^{AB}$, which represents the average electrostatic repulsion between an electron on atom $A$ (in any orbital) and the electron on atom $B$ (in any orbital) and $Z_B$ is the core charge of B (nuclear point charge minus inner shell electrons).

**Two-center one-electron integral $H_{\mu\nu}$ (resonance integral)**

In the MNDO method the resonance integral $H_{\mu\nu}$ is proportional to the overlap integral $S_{\mu\nu}$:

$$H_{\mu\nu} = S_{\mu\nu}\frac{(\beta_\mu + \beta_\nu)}{2}, \quad (7)$$

where $\beta_\mu$ and $\beta_\nu$ are adjustable parameter characteristics of atomic orbital $\phi_\mu$ on atom $A$ and $\phi_\nu$ on atom $B$. For a given first-row or second-row element, there are at most two different $\beta$ parameters, i.e., $\beta_s^A$ for the $s$-orbital and $\beta_p^A$ for the $p$-orbital of atom $A$.

**One-center two-electron integrals**

These integrals in the MNDO method are derived from experimental data on isolated atoms. For each atom there is a maximum of five one-center two-electron integrals, that is ($ss|ss$), ($ss|pp$), ($sp|sp$), ($pp|pp$), where $p$ and $p'$ are two different $p$-type atomic orbitals. The extra one-center two-electron integral, ($pp'|pp'$) is related to two of the other integrals by

$$(pp'|pp')=1/2[(pp|pp)-(pp|p'p')]. \quad (8)$$

If the five independent one-center two-electron integrals are expressed by symbols $G_{ss}$, $G_{sp}$, etc., the Fock matrix element contributions from the one-center two-electron integrals are

$$\begin{aligned} F_{ss}^{\alpha} &= P_{ss}^{\beta}G_{ss} + (P_{px}^{\alpha+\beta} + P_{py}^{\alpha+\beta} + P_{pz}^{\alpha+\beta})G_{sp} \\ &\quad - (P_{px}^{\alpha} + P_{py}^{\alpha} + P_{pz}^{\alpha})H_{sp}, \\ F_{sp}^{\alpha} &= P_{sp}^{\alpha+\beta}H_{sp} - P_{sp}^{\alpha}(H_{sp} + G_{sp}), \\ F_{pp}^{\alpha} &= P_{ss}^{\alpha+\beta}G_{sp} - P_{ss}^{\alpha}H_{sp} + P_{pp}^{\beta}G_{pp} + (P_{p'p'}^{\alpha+\beta} + P_{p''p''}^{\alpha+\beta})G_{p2} \\ &\quad - \frac{1}{2}(P_{p'p'}^{\alpha+\beta} + P_{p''p''}^{\alpha+\beta})(G_{pp} - G_{p2}), \\ F_{pp'}^{\alpha} &= P_{pp'}^{\alpha+\beta}(G_{pp} - G_{p2}) - \frac{1}{2}P_{pp'}^{\alpha}(G_{pp} + G_{p2}), \end{aligned}$$

where $\boldsymbol{P}^{\mathrm{T}}\equiv\boldsymbol{P}^{\alpha+\beta}\equiv\boldsymbol{P}^{\alpha}+\boldsymbol{P}^{\beta}$. where $\boldsymbol{P}^{\alpha}$ and $\boldsymbol{P}^{\beta}$ are density matrices for two different sets of spatial molecular orbitals–those that hold electrons with spin up and those that hold electrons with spin down. By reversing the superscripts $\alpha$ and $\beta$ in the above three equations, one can easily get three similar equations for the beta orbitals Fock matrix elements.

**Core-core repulsion integrals**

From electrostatics the core-core repulsion is $E_N(A,B)=Z_AZ_B/R_{AB}$, where $Z_A$ and $Z_B$ are core charges (nuclear point charges minus inner shell electrons). Since the electron-electron and electron-core integrals do not immediately collapse to the form $c/R_{AB}$ (where $R_{AB}$ is the distance between atom $A$ and atom $B$, and $c$ is a constant) for distances greater the van de Waals radii, there would be a net repulsion between two neutral atoms or molecules. To deal with this, core-core repulsion is approximated by $E_N(A,B)=Z_AZ_B\gamma^{AB}$. In addition to this term, the decreasing screening of the nucleus by the electrons as the interatomic distance becomes very small must be considered. At very small distances the core-core term should approach to the classic form. To account for this, an additional term is added to the basic core-core repulsion integral as follows:

$$E_N(A,B)=Z_AZ_B\times[\gamma^{AB}+(1/R_{AB}-\gamma^{AB})\exp(-\alpha_{AB}R_{AB})] \quad (9)$$

where $\alpha_{AB}$ is a diatomic parameter. The MNDO approximation to the screening effect is similar to that of other methods but has the functional form

$$E_N(A,B)=Z_AZ_B(s_As_A|s_Bs_B)\times[1+\exp(-\alpha_AR_{AB}) +\exp(-\alpha_BR_{AB})]. \quad (10)$$

The core-core repulsion integrals are different for O-H and N-H interactions, and are

$$E_N(A,H)=Z_AZ_H(s_As_A|s_Hs_H)\times[1+\exp(-\alpha_AR_{AH})/ R_{AH}+\exp(-\alpha_HR_{AH})], \quad (11)$$

where $\alpha$ is an adjustable atomic parameter. The numerical values of $\alpha$ are the same for each element.

The mixed models used in AM1 and PM3 are identical, because these methods are derived based on NDDO. The core Hamiltonian correction due to the interaction of the charges between the quantum mechanics region and the classic region is for both and are on atom $A$, and the average electrostatic repulsion between an electron on atom $A$ (in any orbital) and an electron on atom $B$ (in any orbital), is $\gamma_{AB}=\mu_A\nu_A|s_Bs_B$. The interaction energy between the nuclei in the quantum mechanical region and the charges (including the nuclear charges and electronic charges) is

$$\Delta E_N=\sum\{Z_AZ_B(s_As_A|s_Bs_B)\times(1+\exp(-\alpha_AR_{AB})/R_{AB} +\exp(-\alpha_AR_{AB}))-Z_AQ_B(s_As_A|s_Bs_B)\} \quad (12)$$

where $Q_B$ is the electronic charge of atom B, with

$$(s_As_A|s_Bs_B)=\frac{1}{\left(R_{AB}^2+\frac{1}{2}(1/AM_A+1/AM_B)^2\right)^{1/2}} \quad (13)$$

where AM are the monopole-monopole interaction parameters.

## CALCULATION OF PHYSICAL PARAMETERS FOR BASES AND SINGLE NUCLEOTIDES

Knowledge of the physico-chemical properties of isolated canonical nucleotide bases are important for understanding the nature of nucleotide-nucleotide and protein-nucleic acid recognition (Schneider and Berman, 1995; Zheltovsky *et al*., 1989). In (Govorun *et al*., 1992) the AM1 method was applied to calculate dipole moment, heat of formation and total energy of nucleotide bases.

Our further interest was to provide similar cal-culations also for nucleotides themselves. We conducted such calculations by PM3 method, and the results are presented in Fig.1 for dipole moment, total energy and heat of formation showing that dipole moment values calculated by PM3 method for all nucleotides except guanine are very close to ones calculated by AM1 method (Fig.1a). The total energy for nucleotides and their bases behave similaring, but shifted by −27214.2 kJ/mol (Fig.1b). The heats of formation of nucleotides were all negative and also had similar functional dependence shifted by around

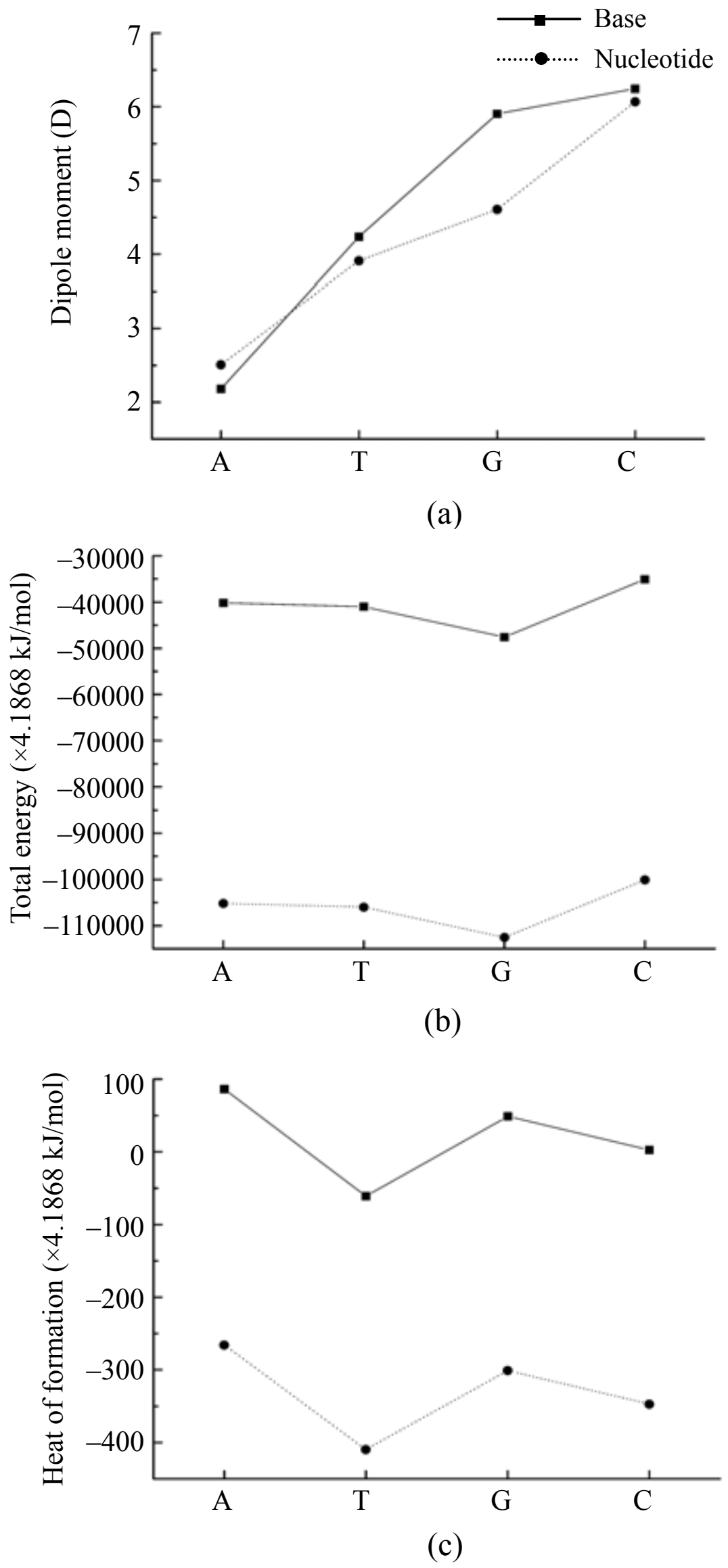

**Fig.1 Dependence of physical properties of isolated nucleotides and their bases from the determinative degree (a) Dipole moment; (b) Energy of most stable formation; (c) Heat of formation**

−1465 kJ/mol (Fig.1c).

Such a difference can be explained by the influence of negative energy of the added disoxiribose atoms and $PO_4$ group.

## DIPOLE MOMENT OF NUCLEOTIDES IN TRIPLETS

The dipole moment of a molecule describes the linear inhomogeneity of its charge distribution and plays an important role in determination of molecule size and can be related with intermolecular interaction in such complex systems as DNA. In the simplest case dipole is a system containing two equal by module but different by sign charges $q$ at the distance $\boldsymbol{r}$, with this system having dipole moment $\boldsymbol{D}=q\boldsymbol{r}$. In the case of liquid medium with irregular charge distribution $\rho(\boldsymbol{r})$ the dipole moment is $\boldsymbol{D}=\int\rho(\boldsymbol{r})\boldsymbol{r}\mathrm{d}^3r$. The dipole may be formed also from one positive or negative charge and zero charge.

We have calculated the dipole moment of all nucleotides taking into account their interaction with their neighbours. In such a way, we obtained the functional dependencies of the dipole moment according to its place in the codon (Fig.2) and to the determinative degree (Fig.3).

Some dipole moments as seen from Fig.2 have the maximum value in the middle point, but sometimes the maximum of nucleotide in one codon differs from its dipole moments in other codons. The nucleotide dipole moments have not so appreciable differences in other two points and some nucleotides have maximum dipole moment in the central point, except some cases of guanine located at the beginning of codons have much lower dipole moment value at two other points.

Let us look at the graphs of dipole moments in more detail. The plot describing dependence of the adenine dipole moment value on its location at the codon shows that all the dipoles moments lie very close to each other when adenine located at the beginning of the codon. If adenine located in the middle of the codon, its dipole moment values were dispersed (Fig.2). In this case adenine had larger value of dipole moment in codons TAG and TAT, than in codons TAA and TAC. In codons AAT and AAA adenine had notably large dipole moment, in codons CAG and CAT it had lower value, but was larger than that in codons CAC and CAA. Also adenine in the codons GAT and AAT (which code the leucine) had larger dipole moment than in other codon coding leucine. The adenine dipole moment in codons GAG and GAC was larger than that in GAA. When the adenine located at the end of the codon, the values of its dipole moments were almost equal except in codons GCA, GGA and AGA, in which the adenine dipole moment was lower than that in others.

The dependence of the thymine dipole moment on its location in the codon (Fig.2) indicated that, if the nucleotide located at the beginning of the codon, its dipole moment's values were almost equal to each other except the case when it in codon TGC, where thymine had lower dipole moment value. When thymine located in the middle part of the codon its dipole moment's values in different codons were dispersed but their dispersal was not as large as those of adenine dispole moment values. Thymine had larger dipole moment in codons TTG and TTC than in codons TTA and TTT. If thymine located at the end of the codon, its dipole moments in different codons were divided into three groups. The largest dipole moment for the nucleotide occured when it located in the first group consisting of codons TGT, CGT, CCT, CTT and CAT. All other codons except ACT were in another group where the dipole moment's value was lower than that in the first group. If thymine loacted in the codon ACT**,** it had the lowest dipole moment's value.

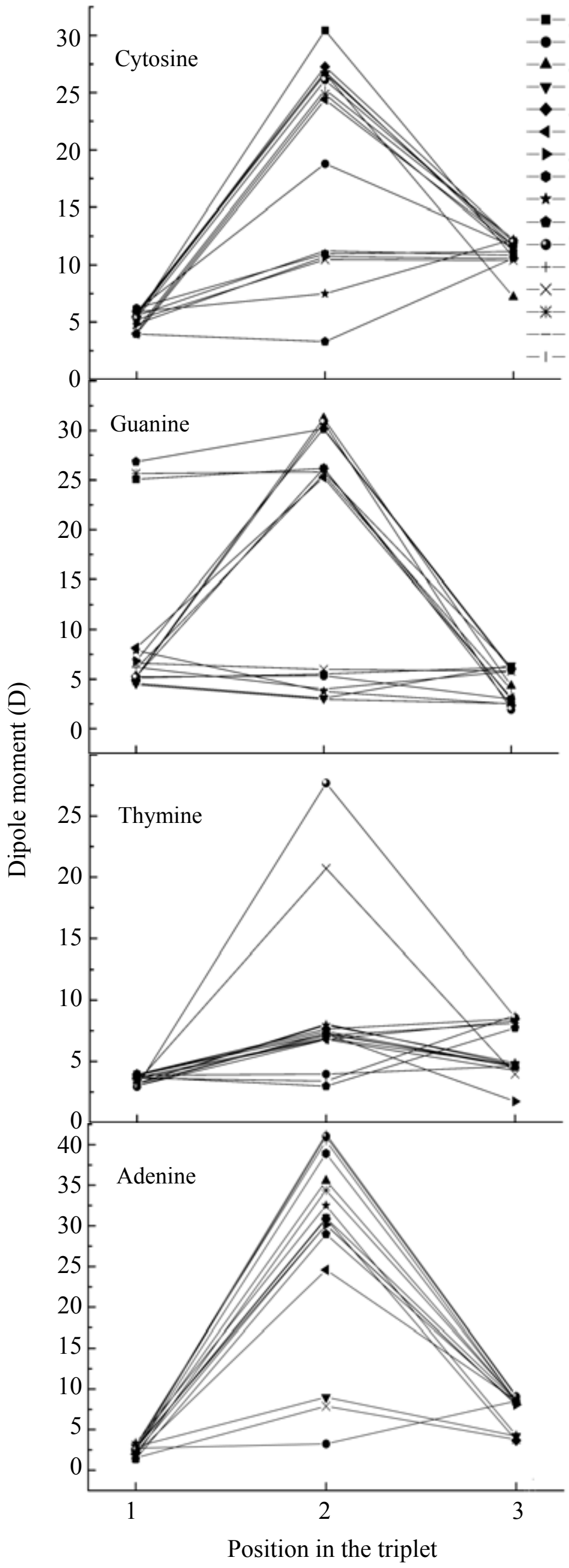

**Fig.2 Dependence of nucleotide dipole moment from its position in codon**

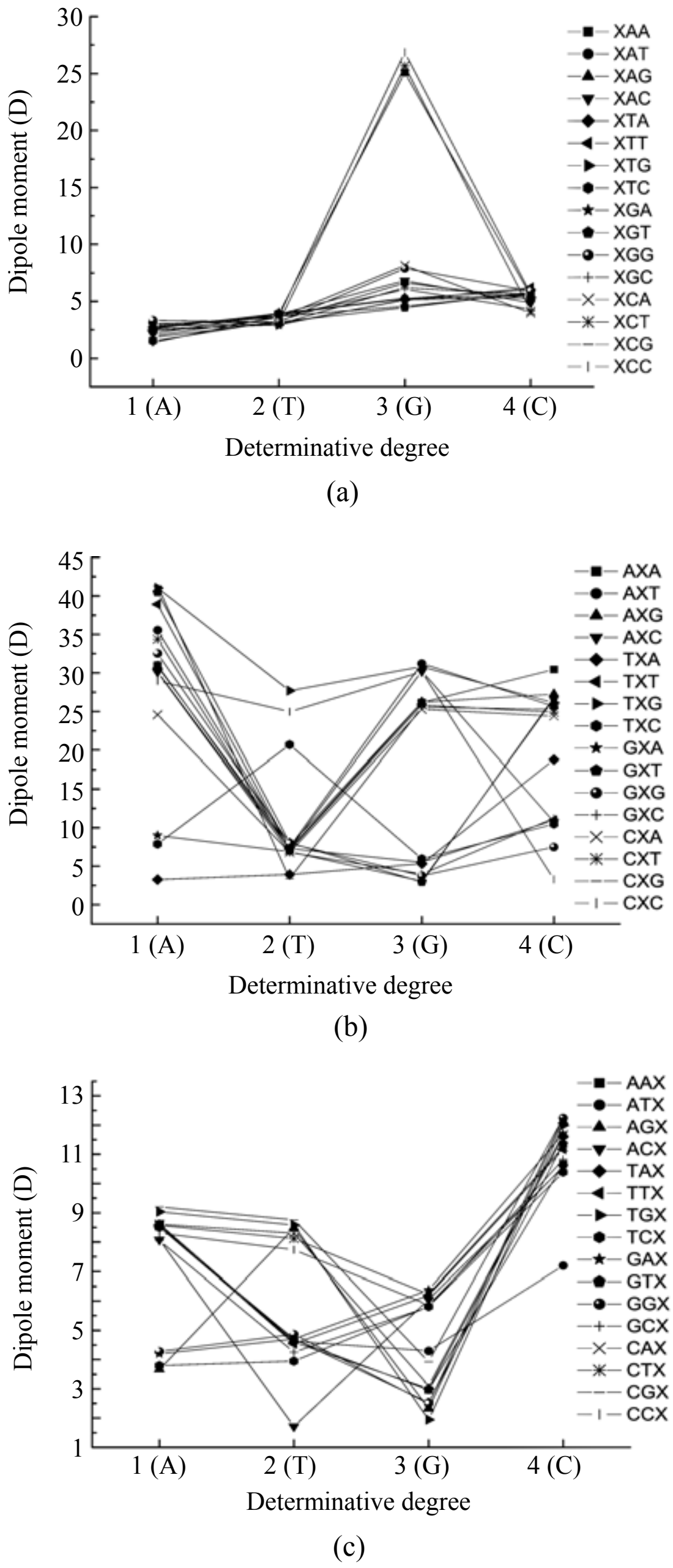

**Fig.3 Dependence of dipole moment of nucleotides based on the different places (X) in codon from determinative degree (a) The first place; (b) The second place; (c) The third place**

The dependence of guanine dipole moment value on its location in the codon (Fig.2) showed that when guanine located at the beginning of the codon it had in codons GCC, GCT and GAA much larger dipole moment than in other codons. When the guanine located in the middle of the codon it had the largest dipole moment in codons AGT, TGG and CGC, the lowest in codons TGC, TGT, TGA, GGC and GGT. If located in the middle of another codons, guanine had medium value dipole moment. When guanine located at the end of a codon, its dipole moments in different codons are almost equal except in the codons GCG, CGG and ATG where the guanine dipole moment is larger than in other codons.

The dependence cytosine dipole moment value on its location in the codon (Fig.2) indicated that when cytosine located at the beginning of a codon it had in different codons almost equal dipole moment except the case when it located in the codons CCC and ACA (cytosine had larger dipole moment in these codons than in other ones). When cytosine located in the middle of a codon it had largest dipole moment in codon ACA, lower dipole moment in codons ACG, GCT, TCG, ACT, CCT and CCA, the dipole moment in the codon TCA was noticeably lower than that in other codons except codons GCG, CCC and CGG, where cytosine had medium dipole moments much lower than that in those codon TCA. When cytosine located at the end of a codon its dipole moments were almost equal except for codon ATC, where cytosine had noticeably lower dipole moment than in other codons.

Fig.3a shows the dependence of dipole moment on the type of nucleotide located at the beginning of a codon. When adenine, cytosine or guanine located at the first position, the values of dipole moment had very little differences, but when guanine was the first nucleotide in the codon the dipole moment values differed a little more (some dipole moments had the maximum at this point). The values of dipole moment of codons GAA, GCT, and GCC were greatly larger than all other codon dipole moment values.

Fig.3b on the dependence of dipole moment on the type of nucleotide located in the middle of a codon shows that when the adenine located in the middle of the codon, the GAT, TAG, TAT and CAG codons had the largest dipole moment values forming a little group, all other codons (except CAA, TAC, TAA and GAA) had a little lower dipole moment values and these values also form a group with little differences from each other. The CAA codon had relatively lower dipole moment value, the codons TAC and GAA had much lower values lying very closely to one another and the TAA codon had the lowest dipole moment value. When thymine located in the middle of a codon the codon TTG had the largest dipole moment value, the CTC codon had lower dipole moment value than codon TTG, the TTC codon had much lower dipole moment value and the codons CTG and TTA had almost equal and the lowest dipole moment values. All other codons had almost equal dipole moment values which were a little larger than the lowest values.

In the case of guanine located in the middle of the codon there were four groups with almost equal dipole moment values individually. The group with the largest values of dipole moment was comprised of AGT, CAC and AGC codons, the group with lowest dipole moment values was comprised of GGT, GGC and GGG codons, the group with a little larger dipole moment value was comprised of TGA, TGT and TGC codons. All other codons formed a group with dipole moment values a little lower than the largest dipole moment value.

When cytosine located at the second position in the triplet, the CCC codon had the lowest dipole moment value, the GCG codon had larger value, the ACC, TCT, TCC and GCC codons had almost equal dipole moment values and that were larger that of codon GCG, the TCA codon had dipole moment value appreciably larger than those of codons in the previous group and the ACA codon had the largest dipole moment value. All other codons had dipole moment values very close to each other and formed a group with dipole moment value in-between that of the largest value and the value of the TCA codon.

Fig.3c shows the dependence of dipole moment on the type of nucleotide located at the third position in the codon. When this nucleotide was adenine the TCA and AGA codons had almost equal dipole moment values, the codons GAA and GGA also had almost equal dipole moment values but they were a little larger than the values for codons TCA and AGA. All other codons values comprised a group including three subgroups with little differences from one another. The subgroup with the largest values comprised two groups with almost equal values corresponding to the codons TGA and CGA, the subgroup with lowest values comprised another two almost equal values groups corresponding to the ACA and GCA, all other codons had dipole moment values in-between the values of the subgroup which were almost equal. When thymine located at the third place in the codon all the codons (except ACT with appreciable lowest value) there were two groups with very little different dipole moment values. The group with largest dipole moment values was comprised of AGT, TGT, CGT, CAT, CTT and CCT codons, all other codons comprised another group.

When guanine located at the third place in the codon, codon TGG had the lowest dipole moment value, codon AGG had a little larger value, the codons GGG and GTG had almost equal dipole moment values which were a little larger than that of codon AGG, the codons AAG, TAG and CAG also had almost equal dipole moment value which were a little larger than that of two previous codons. The codons ATG and CGG had appreciably larger dipole moment values with very little differences. All other codons formed a group with very little differences in dipole moment values. When cytosine located at the third place in the codon all the codons except ATC had very large dipole moment values lying very close to one another. The ATC codon in this case had appreciably lower dipole moment value.

## FORMATION HEAT OF NUCLEOTIDES IN TRIPLETS

The heat (enthalpy) of formation is calculated by substructing atomic heats (enthalpies) of formation from the binding energy, and it is the value that could be more useful than the directly calculated binding energy. Knowing the enthalpy and temperature it is possible to estimate such system's parameter as heat capacity. Usually, temperature is measured more often than the entropy, and the system pressure is more possible to be constant, than the volume. That is why it is more reliable to find the functions for which the measuring parameters are the "natural" variables, such as the heat of formation (enthalpy).

Heat of formation is the energy released during the process of molecule formation by all atoms including it. When the system volume remains constant, then the quantity of heat provided by system $Q$ is equal to its energy change, i.e., $dQ=dE$. If the system pressure also remains constant the change of heat is $dQ=d(E+PV)=dW$, where $W=E+PV$ is the function of heat−the enthalpy.

Here we calculated the heat of formation of all nucleotides taking into account their interaction with their neighbours. In such a way we obtained the functional dependencies of the heat of formation based on the location in the codon (Fig.4) and the determinative degree (Fig.5).

The plots showing energy and heat of formation behavior are similar for every nucleotide and these parameters values at one point have comparatively small difference (Fig.4). These parameters having minimum value at the middle point (2) reflects that the central nucleotide in the codon is more stable than those at the edges.

Fig.5a shows dependence of heat of formation on the type of nucleotide located at the first position in the triplet.

If adenine located at the first place in the codon there are five groups with a little difference in the values of heat of formation. The largest value occurs in the group including the codons ATT and ATC, the AAC codon forms another group which has a little lower heat of formation value. The group with lower heat of formation value includes AAA, AAT, ATG and other codons which do not belong to other groups. In the fourth group there is only AAG codon, the last group which has the least energy among all the groups including AGA and ACC codons.

When thymine located at the first place in the triplet there were multiple groups with different heat of formation values. These groups contain the following sorted by decreasing heat of formation codons: TTT, TCC, TTG, TCA, TGC, TCT, TTC, TGT, TCG, TTA, TAC, TAT, TAA, TGG, TAG, TGA.

When guanine located at the first place in the codon, the codon GTG had minimum heat of formation value at this point. The GTT and GTA codons had maximal heat of formation value, the codons GCC, GAT, GAC, GAA, GCG, GGC and others formed a group with almost equal heat of formation values. The codons GAG, GGT and GCA had lower and almost equal values, the codons GGG, GCT and GCT in the graph are lying one under another, so they are sorted by decreasing heat of formation values.

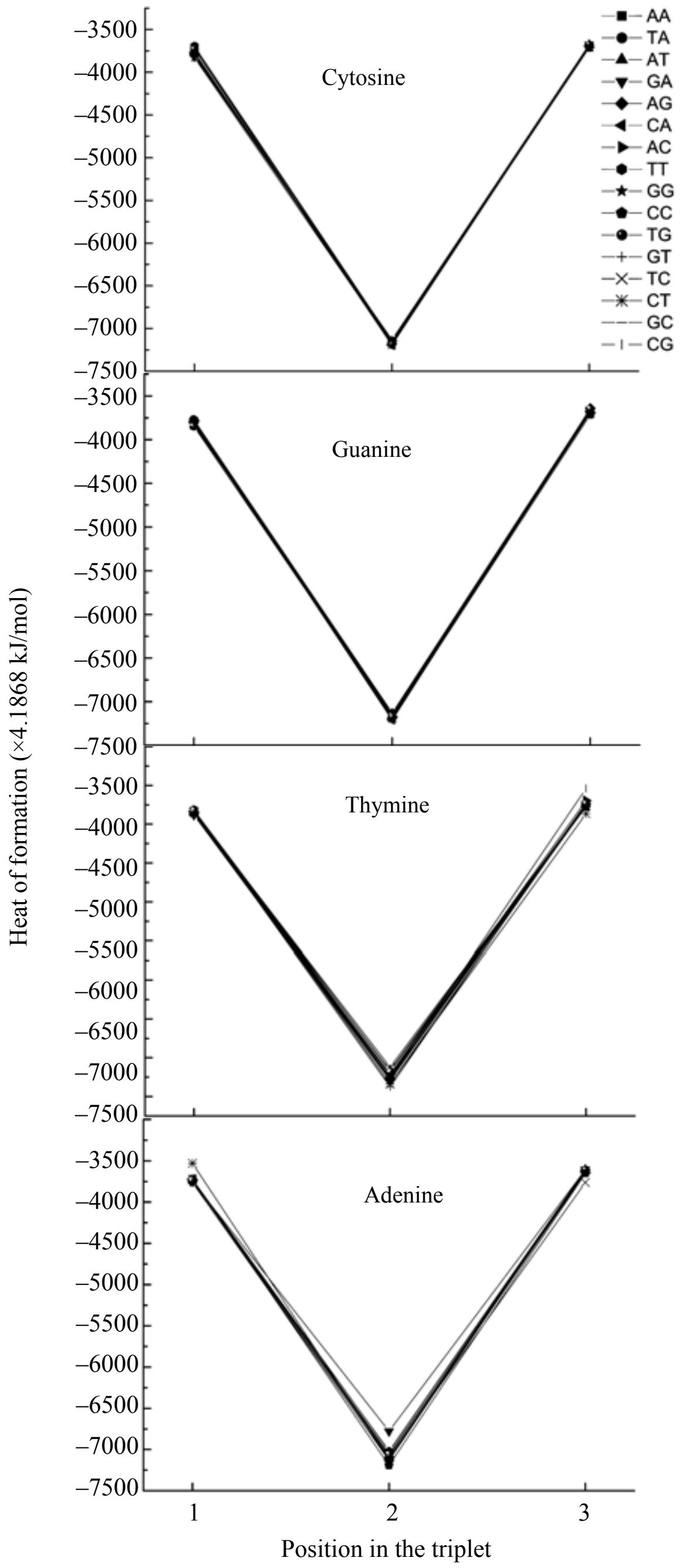

**Fig.4 Dependence of nucleotide heat of formation from its place in codon**

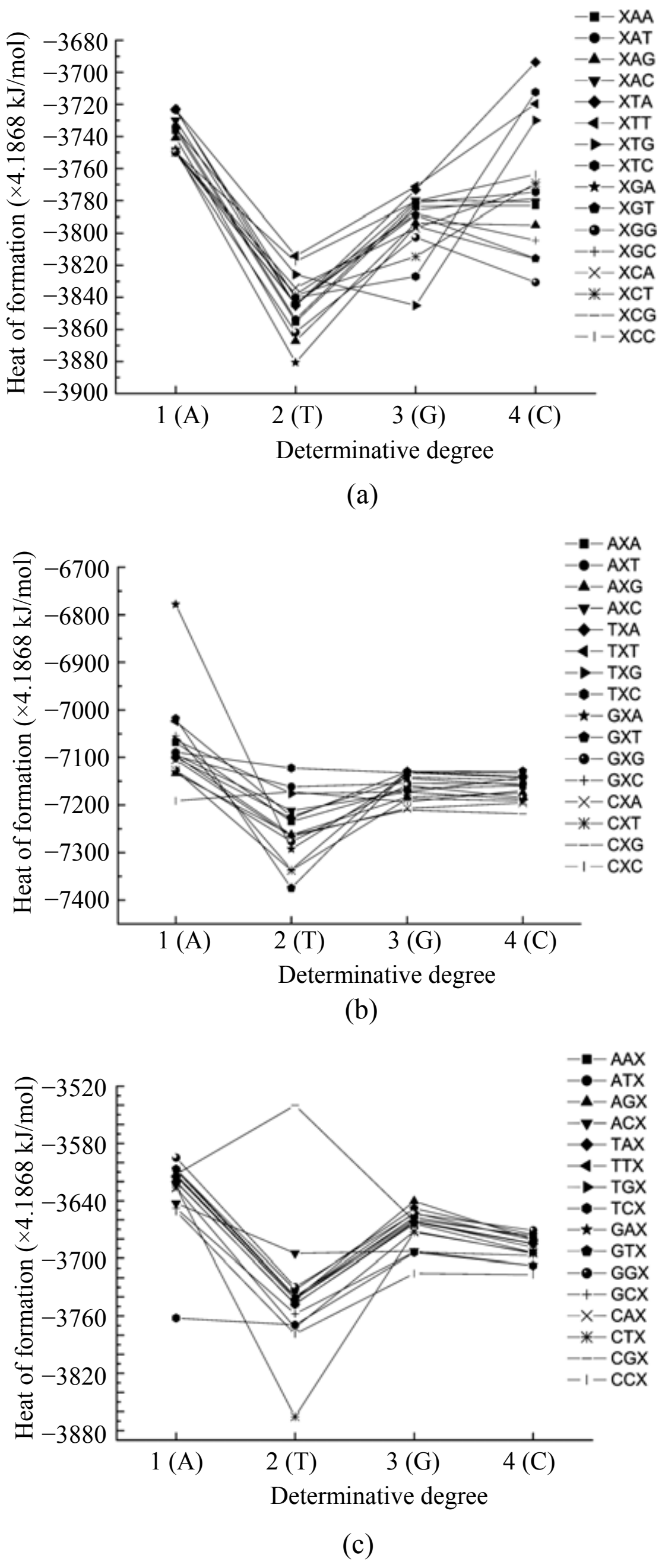

**Fig.5 Dependence of heat of formation for nucleotides on the different places (X) in codon from determinative degree (a) The first place; (b) The second place; (c) The third place**

In the case of cytosine located at the beginning of the codon, the codons CTA, CTC, CTT and CTG had the largest values of heat of formation at this point (the CTA codon had the largest heat of formation value), the codons CAG, CGC, CGT, CGG hadthe lowest values of this parameter (the CGC codon had the minimum value) and all other codons combine into a group with larger value of heat of formation.

Fig.5b shows the dependence of values of heat of formation on the type of nucleotide located in the middle of the codon. If the nucleotide in the middle of the codon was adenine, the codon GAA had appreciably larger energy of heat transformation than all other codons at this point. The codons TAT and GAT form another codon group with lower energy of heat transformation, another group with much lower value of energy formation were the codons GAC, TAG and AAA. The group with the lowest energy of heat transformation contained only CAC codon, the group with a little larger energy of heat transformation contained AAG, CAT and GAG codons. The group containing all other codons with heat of formation values in the graph is the third one from the bottom if the groups are sorted by increasing heat of formation value.

When thymine located in the middle of the codon, the largest value of heat of formation was that of the TTC codon. The group with lower heat of formation values contained the ATT, CTC and TTG codons, the group with much lower heat of formation values contained the following codons: ATC, TTT, TTA and ATA. The group with the least heat of formation value contained the codons GTC and CTT. The group containing all other codons has bigger value than the group described before, but lower value than other groups.

When the guanine or cytosine located in the middle of the codon the heat of formation values of different codons lie very close to one another. In the case of guanine located on the second place in the codon, the GGT, GGC, TGA and TGC codons formed the codons group having the largest heat of formation value. The group with lower values contains the GGA and TGT codons, another group with much lower values contains the AGT and GGG codons.

The next group with much lower heat of transformation value contained the AGA, TGG and AGC codons. The CGA and CGG codons formed the group with the least heat of formation value and the codons AGG, CGT and CGC formed another group with larger values than the group with CGA and CGG codons. If the cytosine located at the second place in the codon, the group with the largest values of heat of formation contained the GCT, GCC, TCA, TCC and GCA codons, the group with lower values contained the GCA, TCT and ACT codons, all other codons except CCG formed the group with much lower values. The codon CCG had the least heat of formation value at this point.

Fig.5c shows the dependence of the heat of formation values on the type of nucleotides located at the third place in the codon. When adenine located at the end of the codon, the GGA codon had the largest heat of formation value, the TCA codon had the lowest value and was appreciably lower than other heat of transformation values, the codons ACA and CCA had larger value than TCA but lower than that of the group containing all other codons forming the fourth group at this point.

When thymine located at the third place in the codon, the codon CGT had the largest heat of formation value, and was appreciably larger than other values, the CTT codon had the least heat of formation value, and was considerably lower than other values. The codons CCT, CAT, TCT and CCT were sorted by increasing heat of formation values, all other codons except ACT formed the group with larger values than those of the four codons described before, the ACT codon had larger value then this group.

When the guanine located at the third place in the codon, the codon CCG had the least heat of formation value, the group with larger values contained the codons GCG, ACG and TCG. All other codons formed the group with the largest values of heat of formation.

In the case when cytosine located at the end of the codon, the codon CCC had the least heat of formation value, the TCC codon had a little larger value and all other codons were in the group with the highest heat of formation value.

ENERGY OF MOST STABLE MOLECULE CONFORMATION IN TRIPLETS

The energy of the most stable molecule conformation is the minimal of its possible energy. The molecule conformation is more stable when this energy is lower. Knowing the value of the most stable conformation energy of the molecules, we can possibly compare which molecule in the set is the most stable. We can do so in the case of molecule parts combined with each other in different sequence. When a part surrounded by other parts has the minimal total energy value the unit containing all the parts in certain order is more stable than any unit containing these parts in another order.

Here we calculated the energy of the most stable conformation of all nucleotides taking into account their interaction with their neighbours. In such a way we obtained the functional dependencies of the energyof the most stable conformation on the basis of the place in the codon (Fig.6) and the determinative degree (Fig.7).

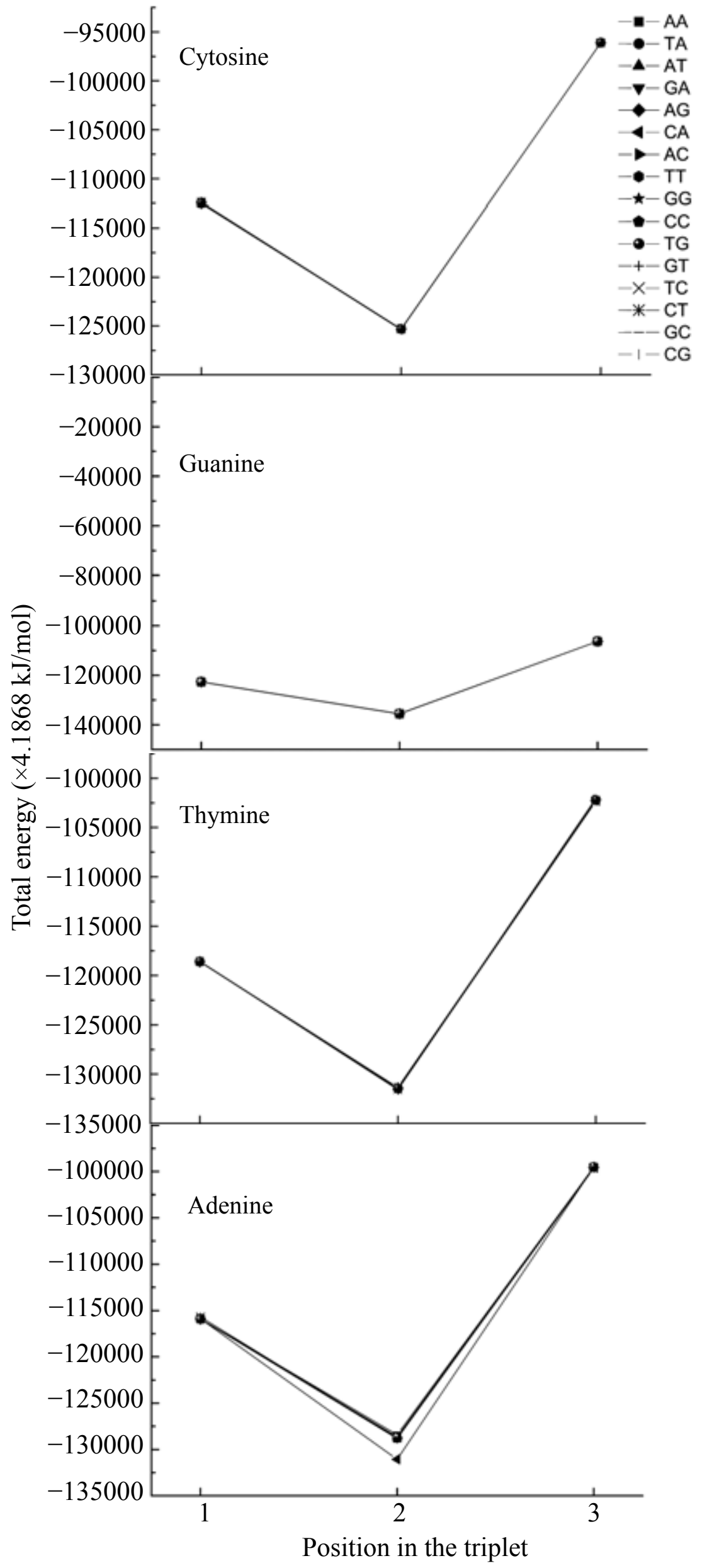

Fig.6 Dependence of energy of most stable nucleotide formation from nucleotides place in codon

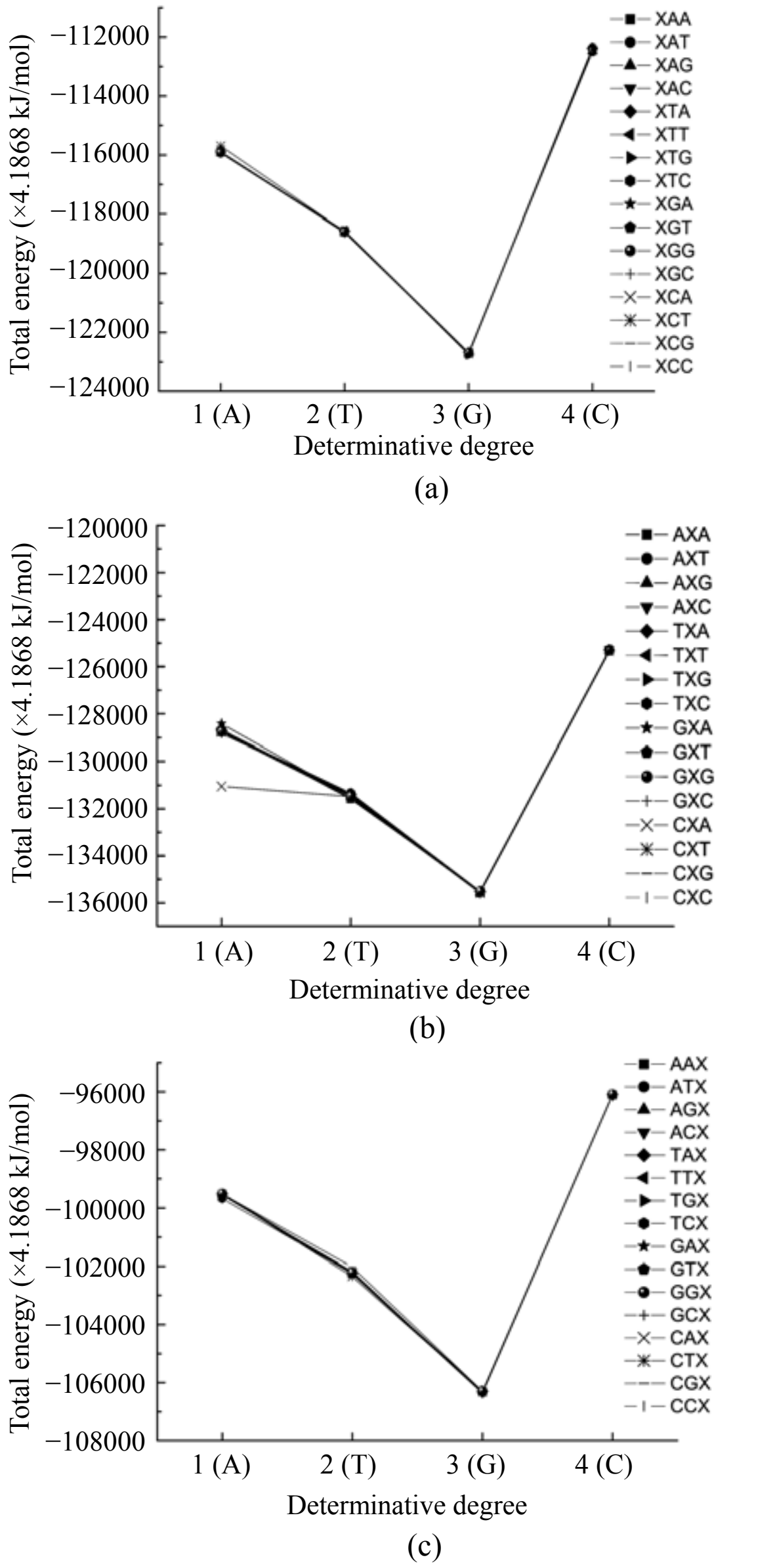

Fig.7 Dependence of energy of most stable nucleotide formation of nucleotides basing on the different places (X) in codon from determinative degree (a) The first place; (b) The second place; (c) The third place

Fig.6 shows that all curves practically coincide except for that of adenine in the middle of the codon. The common trend was getting minimal value at the second position for all cases.

The dependence of the energy of the most stable conformation from the determinative degree (Fig.7) indicated the minimum for the determinative degree $d=3$ (guanine).

COMPARISON OF CODONS AND ANTICODONS

It is interesting to compare physical values of codons and anticodons. In our case we call codon a definite triplet, and its anticodon another triplet having bases complementary to the first one. For convenience, we divide all triplets into two sets, codons

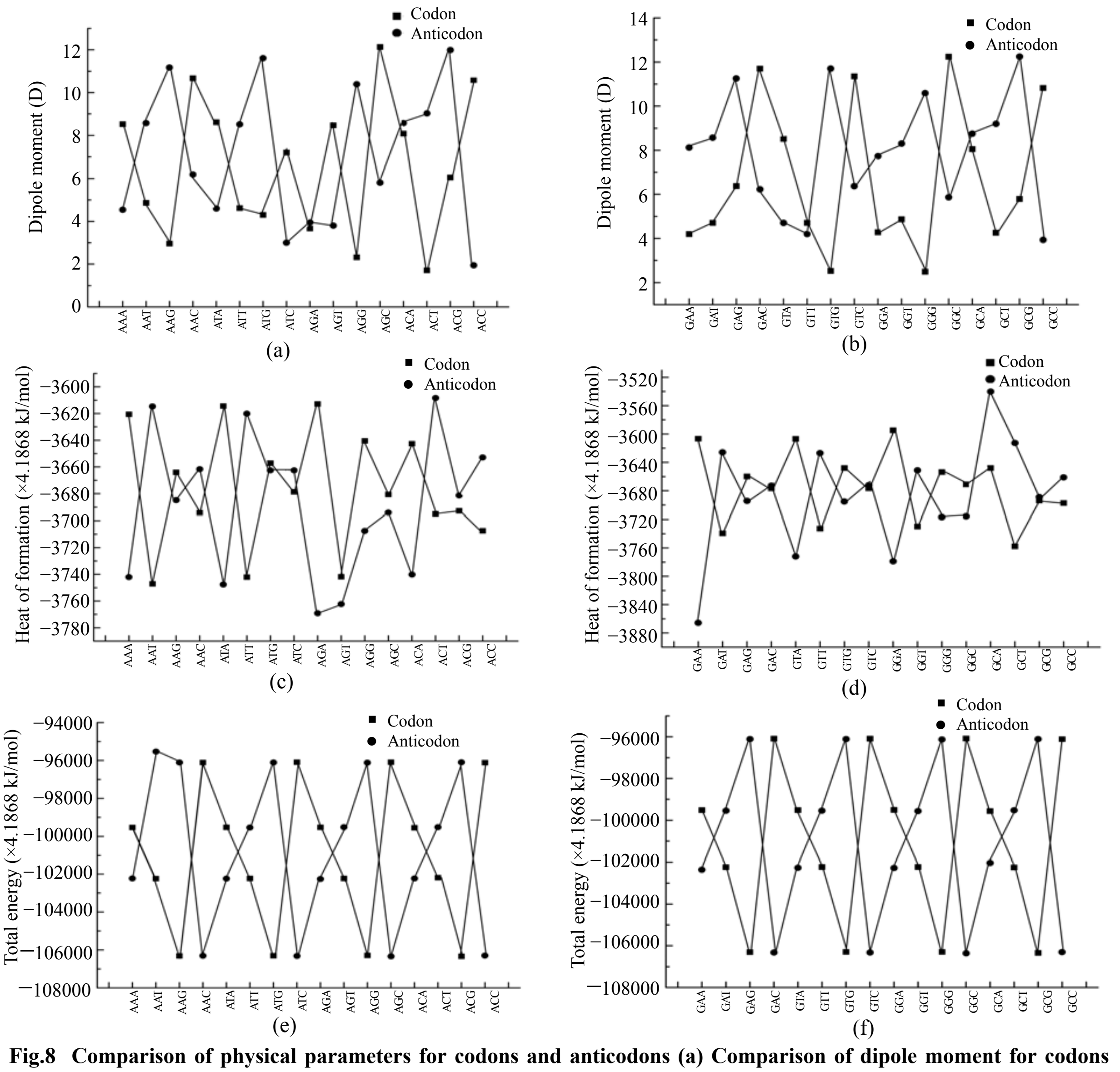

**Fig.8 Comparison of physical parameters for codons and anticodons (a) Comparison of dipole moment for codons beginning from guanine and anticodons beginning from cytosine; (b) Comparison of dipole moment for codons beginning from adenine and anticodons beginning from thymine; (c) Comparison of heat of formation for codons beginning from guanine and anticodons beginning from cytosine; (d) Comparison of heat of formation for codons beginning from adenine and anticodons beginning from thymine; (e) Comparison of energy of most stable formation for codons beginning from guanine and anticodons beginning from cytosine; (f) Comparison of energy of most stable formation for codons beginning from adenine and anticodons beginning from thymine**

and anticodons, by the following rule: codons have A, G on the first place, anticodons have C, T on the first place.

To compare physical values of codons and anticodons, we calculated the dipole moment (Figs.8a and 8b), heat of formation (Figs.8c and 8d), energy of the most stable conformation (Figs.8e and 8f).

When the codon has larg energy value the anticodon has enough lower energy value but when the energy of anticodon is big and the energy of codon is low. Sometimes the energy values for codon and anticodon are almost equal. In this case codon and anticodon have total energy value almost equal to half of their largest energy value. Hence we could propose that in sum the energy values for codon and anticodon are equal to a certain energy value which is a little larger than the largest value for codon or anticodon.

## CONCLUSION

Thus, we have calculated various physical properties (dipole moment, heat of formation and energy of the most stable conformation) for all nucleotides taking into account their interaction with their nucleotides neighbours. This allowed us to obtain special functional dependencies of these variables from their location in the codon and the determinative degree.

## ACKNOWLEDGEMENTS

Authors are grateful to Diana Duplij and Yurij Shckorbatov for their very much fruitful discussions and J. Bashford, D.M. Hovorun, N. Tidjani, C. Zhang for kind sending us their interesting papers. We are thankful to T.L. Kudryashova for language checking.